\documentclass[twocolumn]{aa} 
\usepackage{graphicx}
\usepackage{txfonts}
\begin{document}

\title{High resolution imaging with Fresnel interferometric arrays: suitability for exoplanet detection}

 \author {   L.Koechlin \inst{1}
 	              \and D.Serre  \inst{1}
	              \and P.Duchon \inst{2}     }

   \institute { Observatoire Midi Pyr\'en\'ees, 14 avenue Edouard Belin, 31400 Toulouse, France
         \and
            	Centre National d'\'etudes spatiales, 18 avenue Edouard Belin, 31055 Toulouse Cedex, France  }


\abstract{
We propose a new kind of interferometric array that yields images of high dynamic range and large field. The numerous individual apertures in this array form a pattern related to a Fresnel zone plate. 
This array can be used for astrophysical imaging over a broad spectral bandwidth spanning from the U.V. (50 nanometers) to the I.R. (20 $\mu$m). 
Due to the long focal lengths involved, this instrument requires formation-flying of two space borne vessels. 
We present the concept and study the S/N ratio in different situations, then apply these results to probe the suitability of this concept to detect exoplanets.

}
  \maketitle
  

\section {introduction}

	Images of high angular resolution and high dynamic range are required for the new fields of astrophysics such as exoplanet detection and cartography of stellar photospheres. Multi-aperture interferometry has been used for many years with increasing success. However, in the visible domain, interferometric arrays are still limited to small numbers of apertures: the maximum today being eight,  at VLTI. This relatively modest number of apertures limits the field-resolution ratio in reconstructed images, according to a theoretical limit based on the Shannon theory of information  (Koechlin, Perez  2002). 
	
	To improve the imaging capabilities of interferometric arrays, we propose a setup allowing the recombination of a very large number of beams 
 from very "basic" apertures rectangular holes.
	.The layout of these apertures acts as a diffractive Fresnel plate and directly focus the light (combines the beams) into a point spread function (PSF) of high dynamic range without the need for any reflective or refractive element in the apertures. The focal length of such a Fresnel array can vary from 200 meters to 20 km, depending on the array type and wavelength used. This implies space-borne instruments and formation flying.
		
	Imaging diffractive Fresnel plates have been proposed by Soret (Soret, 1875) and widely used since (Lipson, Lipson \& Tannhauser, 1995).  Large zone plates have also been proposed for space borne instruments, using phase or amplitude modulation of the incident wavefront to focus light in various wavelength domains: submillimetric, I.R.,  visible (Massonnet, 2003), X - and gamma-rays (Skinner 2003). 

	The interferometric setup proposed here for imaging can be regarded either as particular kind of Fresnel plate, or as an aperture synthesis array with a very large number of apertures: thousands to hundreds of thousands. The apertures are rectangles punched into a thin metal foil framing the array. The array when unfolded in space may have a span of a few to a few hundred meters (Koechlin 2004). Other interferometer concepts involving  large numbers of apertures have been developed in recent years, such as Carlina (Labeyrie et al, 2004), but with classical telescopes forming the individual apertures.
	
	Fresnel zone plates are chromatic, and so is a Fresnel array. For zone plates, the chromaticity issue has been addressed and a correction scheme proposed by Falklis \& Morris  (1989). For Fresnel arrays, a similar chromatic correction in the focal instrument provides an achromatic image for spectral bandwidths up to ${\Delta \lambda \over \lambda} = 0.15$
 as discussed in section~\ref{achromatizer}
. The same primary array can be used over a wide range of spectral bands whose central wavelength is tuned by varying the position of the focal satellite along the optical axis of the array. The physical properties of the foil defining the primary array limit the observable domain to a "global" bandwidth spanning from 50 nm in the U.V. to 20 $\mu$m in the I.R.. The limit  at short wavelengths is the transparency to UV radiation of the thin metal foil defining the aperture edges. Towards long wavelengths, the limit is set by thermal radiation from the foil itself at the temperature to which it can be passively cooled: 40 to 120 K, depending on the baffling.
	
In the following sections, we present the design of an interferometric Fresnel array, the percentage of the light going into the Point Spread Function, the dynamic range and how it can be improved by apodization,  the effects of aberrations, the achromatizer design and  Exoplanet detection simulations.


\section {Design of the interferometric Fresnel array   \label{design} }

We propose a design where aperture edges follow only two orthogonal directions, the set of apertures (the array) forming a large mask. The transmission law $T(x, y)$ of this mask can be built as follows. Let us first define functions $g$ and $h$ as:
$$  g(a) = 1  \;  {\rm if} \sqrt {a^2+f^2} \in \left[  \left( k+ {f \over {m \lambda}} +{1 \over 2} \right) m \lambda  \, ;  \left( k+{f \over { m \lambda}} +1\right) m \lambda \right[  $$
$$ { \rm and}  \;  g(a) = 0 \;   { \rm otherwise}  \, ,  $$
$$ h = 1-g$$
where $a$ is the distance from optical axis,
$m$ is the diffraction order ($m=1$ in our application),
$k$ a variable integer: the Fresnel zone index and
$f$ the desired focal length for the array.

The function $g$ has an opaque central segment, whereas $h$ has a transmissive central segment. If developed radially,  $g$ or $h$ define circular Fresnel zone plates. In the present case, the transmission law $T(x, y)$ of our 2D array is based on an orthogonal development of $g$ and $h$: 
$$T_c(x,y) = h(x)g(y) + g(x)h(y) $$
for a "closed  central square" array, and  
$$T_o(x,y) = h(x)h(y) + g(x)g(y) $$
for the complementary "open central square "array.

The actual transmission laws depart from $T_c$ or $T_o$, to achieve apodization (and mechanical consistency of the grid) as described later in the paper.
 \begin{figure}
 \includegraphics[width=88mm]{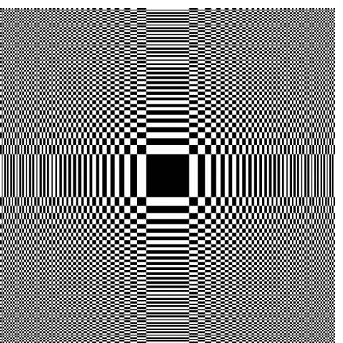} 
   \caption{   Example of $T_C(x, y)$ for a  $k_{max} = 30$ Fresnel zones, i.e. 7200 apertures orthogonal Fresnel array. } 
\label{lentille}
 \end{figure}
Each aperture (open rectangle) in the array can be referenced by its Fresnel zone indices $(k_x$ and $ k_y)$.  
{ 
 The term "Fresnel zone" defines an area delimited in the aperture plane by two concentric circles. These circles are the intersection of the aperture plane with spherical waveplanes centered on the focus and whose radii differ by one wavelength. The central Fresnel zone is the disc delimited by the smallest intersection.
The number of zones covered by a Fresnel array (as for a filled aperture) corresponds to the number of zones crossed from center to edge along a 1D line.
For a square  array of size $C$ and $k_{max}$ Fresnel zones, the  distances between centers of neighboring apertures in the x or y directions are the pseudo periods:
$$p_x = {x \over 2 k_x} = {C \over 4 \sqrt { k_{max} k_x} }  \;  \;  \;   { \rm and}  \; \;  \;  \;   p_y = {y \over 2 k_y} = {C \over 4 \sqrt { k_{max} k_y} }$$
The apertures cover half the pseudo periods, or less if the array is apodized. The focal length of the array is
$$f = C^2/8 k_{max} \lambda$$
and the linear PSF half size:
$$\rho = C/ 8 k_{max}$$
For example a 6 m, 600-zone array has an 8 km focal length at $\lambda = 0.9 {\rm \mu m}$, and a linear PSF half size $\rho = 1.25  {\rm mm}$. 
 \begin{figure}
\includegraphics[width=88mm]{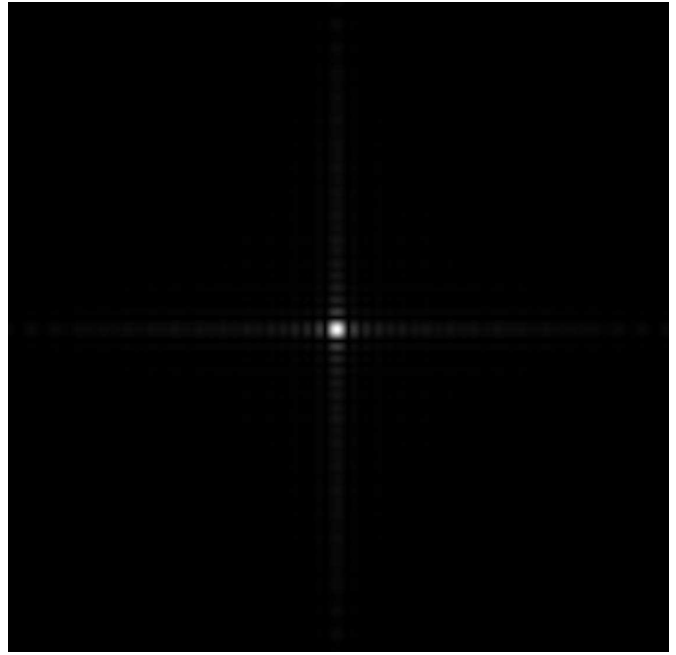} 
\caption{  Computer generated point spread function (PSF) of a 100-zone (center to edge) 80000-aperture Fresnel array. The intensity is displayed at the $1/2^{nd}$ power, to enhance the low luminosity regions of the PSF.}
  \label{psf100}
 \end{figure}

Seen as a diffractive zone plate, this synthesized aperture directly forms images of high dynamic range. Most of the light that escapes from the central part of the PSF is confined to a pair of orthogonal spikes, rather than spread around as for the PSF of circular zone-plates. Orthogonal Fresnel arrays are also adapted for the "apodized square aperture" approach (Nisenson \& Papaliolios, 2001), further improving the dynamic range.

Seen as an interferometric array, this design has the advantage of not requiring any reflective nor refractive element in the primary apertures, and still recombining the beams in a common focus. The principle here is to block the part of wavefront having non-desired phases as seen from the focus. 
\begin{figure}
\includegraphics[width=88mm]{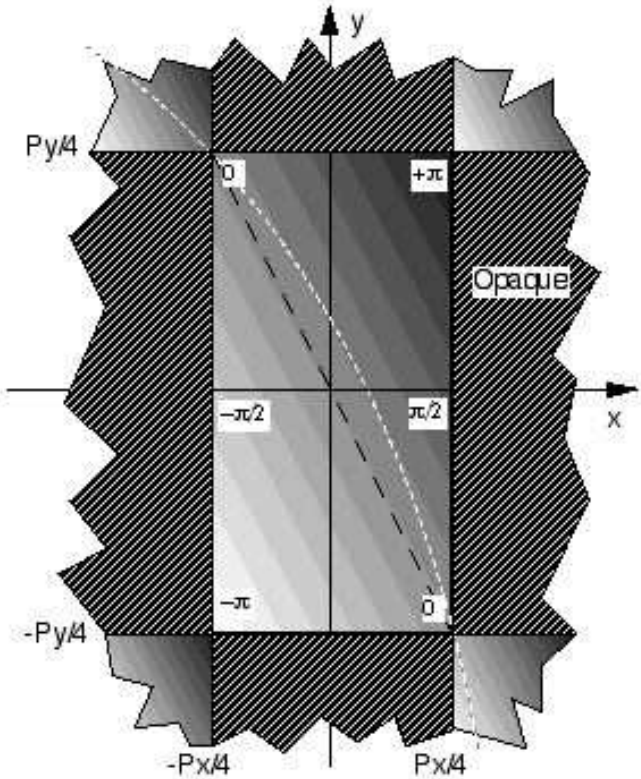}  
\caption{  evolution of the phase within an individual aperture, as seen from common focus.}
\label{fig_integ_motif}
\end{figure}
%


\section { Percentage of the incident light going into the $m^{th}$ order Point Spread Function \label{percentage} }

As Fresnel arrays act by diffraction, only a fraction of incident light is sent into the prime focus.   As for gratings, there are several diffraction orders that get unequal shares, the biggest of them being for order zero. We define the "diffraction efficiency" of a Fresnel array as the proportion of incident light going into the first order focus.

The theoretical diffraction efficiency $E_{circ}$ for a circular Fresnel zone plate when the number of zones tends to infinity  is:
 $$E_{circ} = 1/ \pi^2 m^2  \,\,  {\rm for}  \, m  \, {\rm odd \;  ; \;  } E_{circ} = 0  {\rm \;  for}  \;  m  \; {\rm even \; and \neq 0 }$$ 
where $m$ is the diffraction order.  At the prime focus, which corresponds to the diffraction order $m = 1$, about  10.1\% of the light is gathered.  
  
 In the case of the orthogonal array, the diffraction efficiency can be derived using Cartesian coordinates:
   
  The wave contribution at focus can be expressed:
 $$\Psi_m = A_0 \int  \int_{rectangle} exp (i\phi (x, y)) dx dy.$$

The phase within a given aperture is a quadratic function of position, but, when the number of apertures is large, can be approximated by:
 $ \phi (x, y) = 2\pi ({m x \over p_x} + {m y \over p_y})  .$
 $$ \Psi_m = \int_{-p_x/2} ^{p_x/2}  \int_{-p_y/2} ^{p_y/2} exp (2i\pi {m x \over p_x}) exp (2i\pi {m y \over p_y}) dx dy, $$
 which when $m \ne 0$, yields:
 $$  \Psi_m = A_0 { p_x p_y \over \pi^2 m^2}  \,\,{\rm for} \,m \,{\rm odd \,\,\, ; \,\,\,} 
 \Psi_m = 0  \,{\rm for} \,m\,{\rm even. }$$
For $m$ odd and $\ne 0$, the diffraction efficiency in amplitude for a given element of an orthogonal Fresnel array can be expressed as:
  $$ {\Psi_m \over  \Psi_0} = {4 \over \pi^2 m^2} $$ 
 and considering a 50\% overall "void to total area" transmission ratio , the diffraction efficiency in amplitude is
 $2 / \pi^2 m^2$.
 
 Finally, the diffraction efficiency in intensity  for an orthogonal Fresnel grid is:
$$E_{grid} = {4 \over \pi^4 m^4} .$$
 
At order $m=1$,  this corresponds to a 4.1\% efficiency (when no apodization is applied). 

The effective luminosity of a diffractive square array of size C is the same as that of a reflective circular mirror of diameter $D = 2C \sqrt {E_{grid} \over \pi} = 0.23 C$ on an extended object, and $D  = C \sqrt {1.22} \root 4\of{E_{grid}}= 0.497 C$ on a point source. The angular resolution and dynamic range of  a square array of size C are equal or better than that of a circular mirror of diameter  $D =  C$.

\section { Dynamic range \label{dynamic} }

In order to assess the performance of feasible Fresnel arrays i.e. with a reasonable number of apertures, we have computed PSFs for different test arrays. The dynamic range tests presented in this section (see Fig.~\ref{improve_dynamic_N}) are a comparison between arrays of 125, 250, 500 and 1000 zones (respectively $1.25 10^5$, $5.10^5$,  $2.10^6$ and $8.10^6$ apertures). Figs.~\ref{improve_dynamic_N} and the following show normalized diagonal cuts of the PSF. They are x-labeled in units of resels from the center of the field. A resel, or resolution element has an angular extension of  $\lambda / C$.

We have built these arrays with the transmission law described in section~\ref{design}:   $T_c(x,y)$. The PSF of such arrays is the square modulus of the Fresnel transform of  $T_c(x,y)$, noted as $ \hat{T_c} (u,v)$ in the following. 

As variables x and y are orthogonal, $ \hat{T_c} (u,v)$ can be computed from the functions $ \hat g$ and $\hat h$, which are respectively the Fresnel transforms of the $g$ and $h$ one-dimensional transmission laws (also defined at section~\ref{design}). 
$$ \hat{T_c} (u,v) = \hat h(u)\hat g(v) + \hat g(u) \hat h(v) $$
Relation $T_c(x,y) = h(x)g(y) + g(x)h(y)$ still holds for apodized arrays, however $h \neq (1-g)$ if apodization is applied. 

The fact that two-dimensional arrays can be tested with calculus reduced to 1D for most of the process, greatly improves the computation speed and memory requirements, thus allowing the test of large arrays with little computing power.

As expected, when the number of zones increases, the normalized PSF of a square array tends to that of a filled square aperture of the same size. The numerical simulations show that, in order to achieve a $7. 10^{-6}$ rejection factor (without apodization) at $\simeq$ eight Airy radii from the center, a $k_{max}=1000$-zone array is required. A 250 zone-array will only give $7. 10^{-5}$. Increasing $k_{max}$ has drawbacks, such as decreasing the size of all including the smallest apertures of the array, at a fixed array size. This leads to an increased impact of the element positioning errors on the wavefront quality. Using a complementary pattern ($T_o$ instead of $T_c$ ) leads to a similar performance in the non-apodized case, as one would expect, but causes a modulation reversal in the outer patterns of the PSF. These differences in the outer (low light level) part of the PSF have an impact on the dynamic range. They are due to the fact that Babinet's Theorem is only an approximation in the case of a limited number of Fresnel zones. \\
\begin{figure}
 \includegraphics [width=88mm]{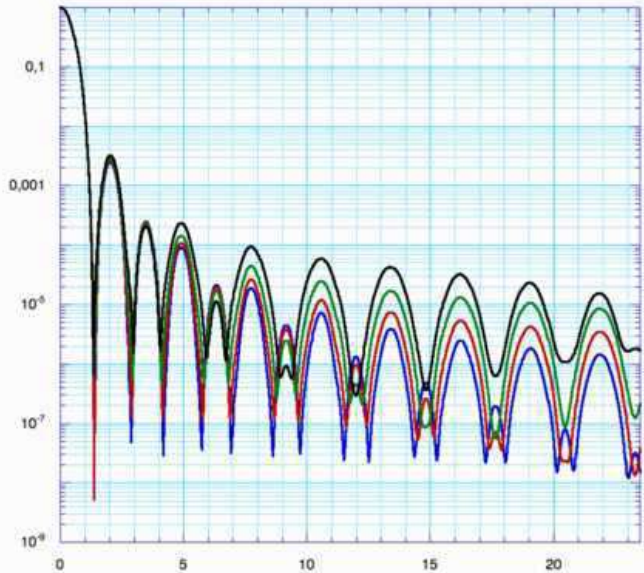}
\caption{  Diagonal cut of the PSF, for  non-apodized arrays having 125, 250, 500 and 1000 zones: respectively 1.25 10¬5, 5.10¬5,  2.10¬6 and 8.10¬6 apertures top to bottom curves. The PSF are computed by Fresnel transform with a wavefront sampling adjusted to 1/25th of the smallest aperture for each array.}
\label{improve_dynamic_N}
\end{figure}

The dynamic range of a given array can be greatly improved by apodization. We have tested different apodization laws (see section~\ref{apod}). We have also tested the effects on the PSF of wave-plane aberrations caused by aperture mispositioning (see section~\ref{aberrations}).

\section { Apodization \label{apod}}

A way to improve the dynamic range of a Fresnel array at a fixed number of apertures is to apodize. An apodized transmission $T_{ap} (x, y)$ is for example:
$$T_{ap} (x, y) = T_c(x, y) Apod(x, y)$$
As the array is orthogonal, the apodization function can be separated into $x$ and $y$:
$$Apod(x, y) = Apod_x (x) Apod_y(y) .$$

Apodization can be done in several ways on a Fresnel array. One of them is  a transmissive or reflective element at a secondary (cooled) pupil plane in the focal setup. Another would be to modulate the aperture positions in order to obtain a phase effect similar to what is proposed by Guyon (2003) for Phase-Induced Amplitude Apodization.

In this paper we present apodization by modulating the apertures areas at the primary array level. 
Apodizing this way somewhat worsens the I.R. noise contribution of the primary array by increasing its radiating surface. However, it simplifies the design of focal instrumentation and provides a very robust means of apodization. Actually, combinations of the different apodizations described above may be used and combined with coronagraphic devices.

The complex amplitude contribution of an aperture is not proportional to its area, as the waves from a given aperture do not all interfere constructively (see Fig.~\ref{fig_integ_motif}). Thus, to fit an apodization law: $Apod (x, y)$, the dimensions of an aperture centered at  $x_c$ and $y_c$ must be:
$$ a(k_x) = {p_x\over \pi} \arcsin Apod_x(x_c)  \;  \rm{and} \;  b(k_y) =  {p_y\over \pi} \arcsin Apod_y(y_c). $$
To assess the gain in dynamic range and loss in transmission for different apodization functions, we have computed the corresponding PSF and global transmission. 
Pending an algebraic derivation of an optimal $ Apod(x, y)$ for Fresnel arrays, akin to the prolate 
apodizing functions for filled square apertures, (Aime, Soummer and Ferrari, 2002), following Papaliolios and Nissenson (2001) and  Soummer (2003), we have evaluated functions of the form:\\
$Apod_{Gauss}(x) = \exp (-x^2/{x_0}^2)$ ; $x_0$ defined such that transmission  becomes $a_0$ at the limb: \\
\indent $ x_0 = - c / 2\sqrt {\log a_0} $\\
$Apod_{cos}(x) = \cos [{2 x\over C} acos(a_0)]$,\\
and $Apod_{cos^2}(x) = { \left[ \cos  \left( {2 x\over C} acos(\sqrt{a_0}) \right)\right]}^{2}$,\\
where $a_0$ is the residual transmission at the edge.
\begin{figure}
 \includegraphics[width=88mm]{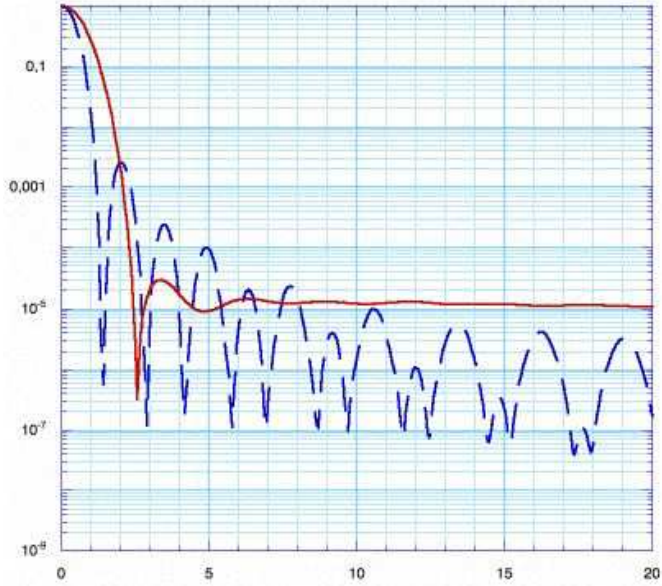}
\caption{ Normalized PSF (diagonal cut) for a 600 zones $T_c$ type array, non apodized (dashed line) compared to apodized with $Apod_{cos^2}(x)$ ($a_0 = 0$, full line). }
\label{apo_left}
\end{figure}
\begin{figure}
  \includegraphics[width=88mm]{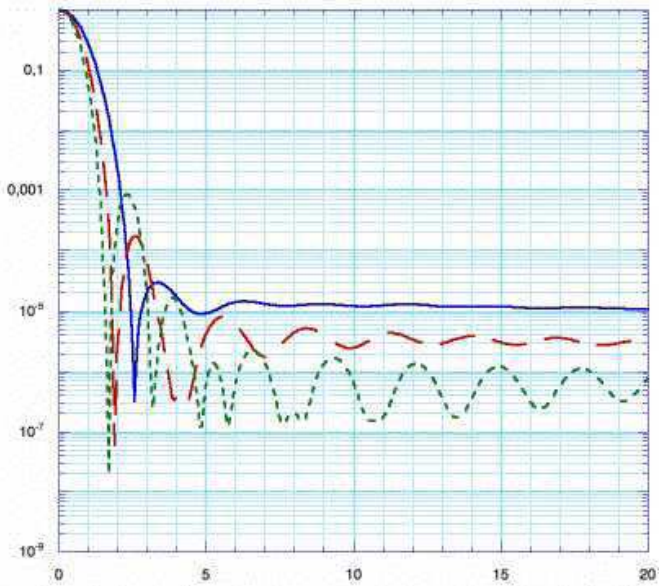}
\caption{ Normalized PSF (diagonal cut) with apodization functions $Apod_{cos^2}$ ($a_0 = 0$), $Apod_{sqrt}$ ($a_0 = 0$), and $Apod_{trig}(x)$ ($a_0 = 0.1$), respectively full, dashed and dotted lines.}
\label{apo_right}
\end{figure}
Apodizing a Fresnel array with $Apod_{cos^2}(x)$ leads to damped variations of the PSF secondary peaks but a high background level, and low overall transmission (see Fig.~\ref{apo_left}  and Table~\ref{tab_transm}).

In order to reach a better rejection factor close to the center of the PSF, we have investigated "broader" apodization functions and apodization starting from $T_o$ instead of  $T_c$ arrays.

Two results compared to previous $Apod_{cos^2}(x)$ are presented Fig.~\ref{apo_right}. The apodizations applied on the wave contribution are:\\
$Apod_{cos^2}(x)$ = $sin({\pi \over2}({2 \over \pi} asin(Apod_{cos^2}(x))))$\\
$Apod_{sqrt}(x)$ = $sin({\pi \over2}{({2 \over \pi} asin(Apod_{cos^2}(x)))}^{1/2})$\\
$Apod_{trig}(x)$ = $sin({\pi \over2}sin({\pi \over2}Apod_{cos^2}(x)))$\\
\\

$Apod_{sqrt}(x)$ 
 leads to "ringing" in the outer parts of the PSF
 when $a_0 \neq 0$, whereas $a_0 = 0$ causes poor transmission. $Apod_{trig}(x)$ with  $a_0 = 0.1$, applied to a $T_o$ type grid, leads to a 2.1\% transmission and a rejection factor better than $3*10^{-6}$ on the diagonal, at 5 resels from center. 
 This last apodization is used when generating the PSF for testing exoplanet detection in the next sections.

\begin{table*}
 \centering
\begin{tabular}{|c|c|c|c|c|c|}
\hline
Type of Fresnel plate & circular & square & \multicolumn{3}{|c|}{square}\\
 \hline
 \hline
apodization & \multicolumn{2}{|c|}{no apodization} & \multicolumn{3}{|c|}{apodization} \\
\hline
Value domain of $Apod$ & \multicolumn{2}{|c|}{ } & \multicolumn{2}{|c|}{1 $\rightarrow$ 0} & 1 $\rightarrow$ 0.1\\
  from center to edge & \multicolumn{2}{|c|}{ } & \multicolumn{2}{|c|}{ } & \\
  \hline
 Applied transmission law & \multicolumn{2}{|c|}{ } & $Apod_{cos^2}$ & $Apod_{sqrt}$ & $Apod_{trig}$\\
\hline
Grid "positivity" & \multicolumn{4}{|c|}{opaque central pattern} & void c. p.\\
\hline
Transmission & 10.1\% & 4.1\% & 0.25\% & 0.90\% & 2.1\%\\
\hline  
\end{tabular}
\caption {Transmission efficiency at focus of diffraction order $m=1$ for different types of diffraction patterns and apodization laws}  
\label{tab_transm}
\end{table*}
\begin{figure}
\includegraphics[width=88mm]{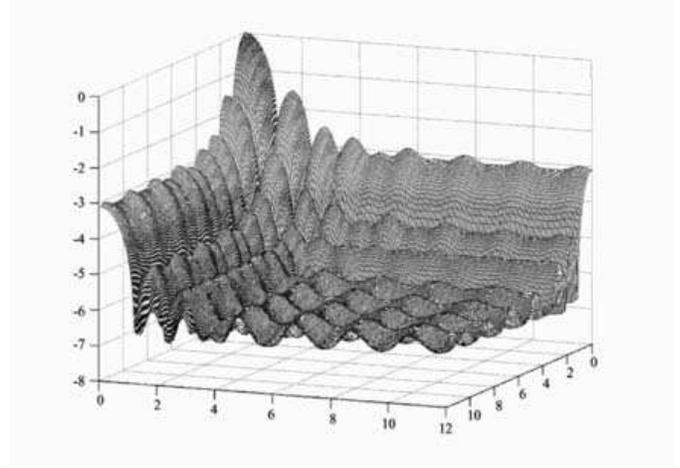}
\caption{  Three-dimensional representation of the PSF resulting from $Apod_{trig}$. Vertical scale is in powers of 10.}
\label{psf_3d}
\end{figure}
 From Fig.~\ref{psf_3d}, one can see that the dynamic range is high in most of the field. In order to achieve that in a whole field, a two-step procedure can be used: an exposure at a given orientation of the PSF, then a second after a 45 degree rotation around the optical axis.
Each exposure is then split into eight sectors:  the four 45 degrees centered on the spikes and the four 45 degrees centered on the diagonals. A composite image is obtained, splicing the four best sectors (diagonals) of each image.

This two-step procedure is not required, except for high dynamic range imaging: the spikes in a single exposure PSF of a Fresnel array contain less energy than those caused by a standard telescope spider.

We have measured PSF residuals in a field defined by the composite image described above. At less than 4.5 resels from the center of the PSF, imaging an exoplanet is not feasible due to the poor rejection factor. From 4.5 to 5.5 resels from the center, rejection is better than $6*10^{-6}$ in $\simeq$ 80\% of the field, so imaging is partly conceivable. At 5.5 resels or more from the center of the PSF, the rejection is everywhere better than $6*10^{-6}$. Beyond 8 resels, the rejection gets better than $2*10^{-6}$.
\begin{figure}
\centering
 \includegraphics[width=55mm]{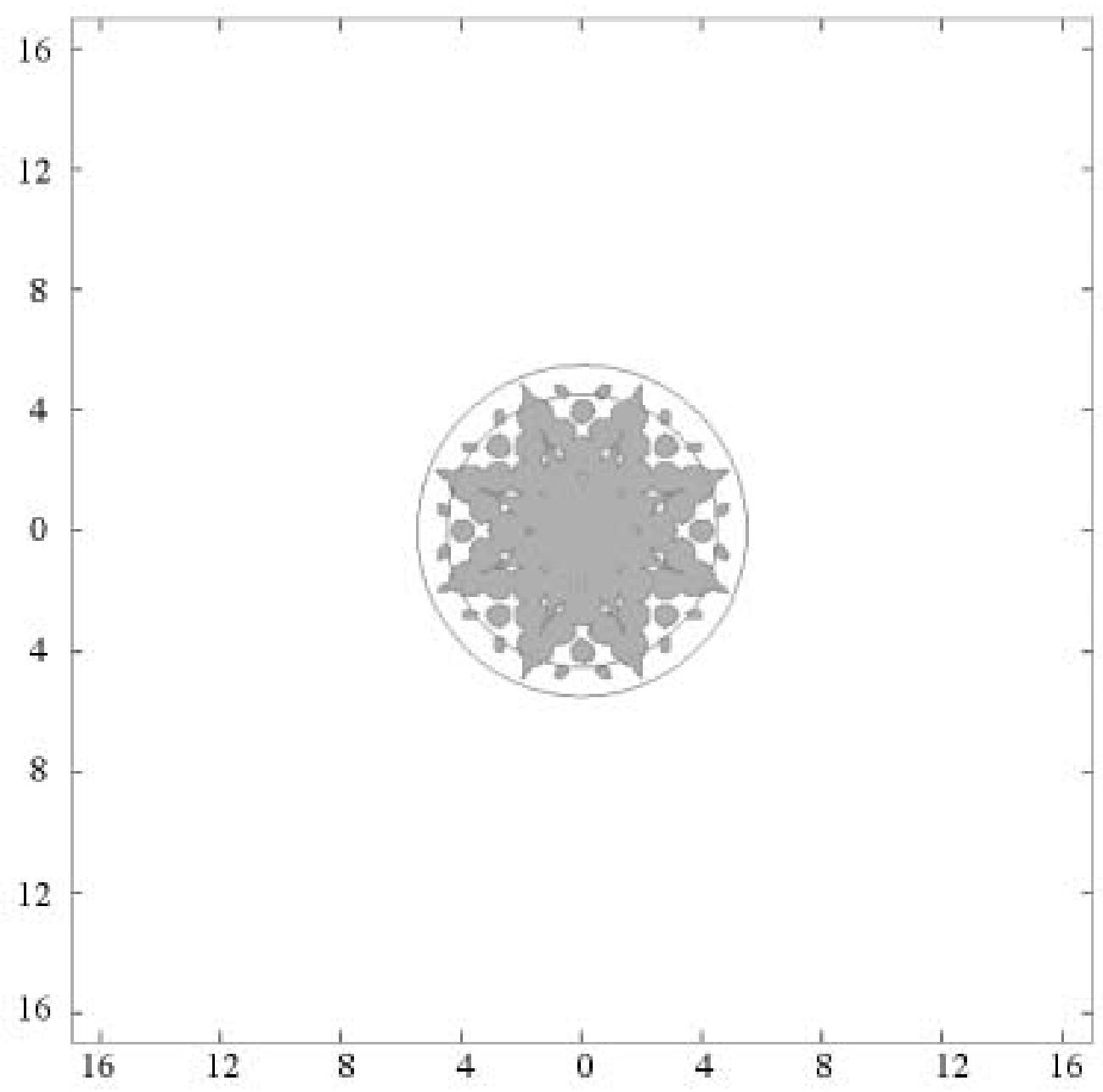} 
 \includegraphics[width=55mm]{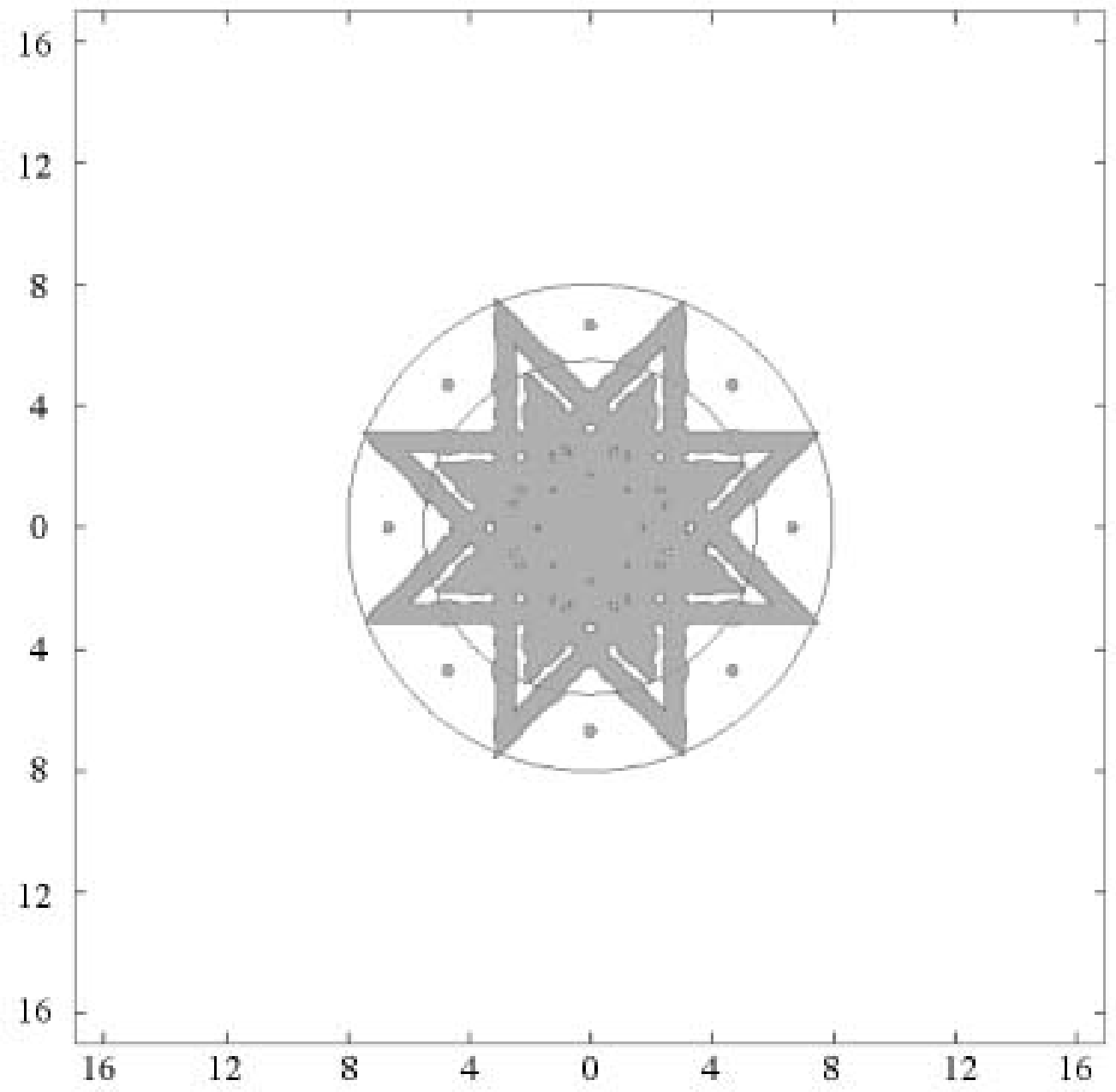}
\caption{  {\bf top:}  Contours of the PSF resulting from the above observing procedure. Shaded areas indicate where the rejection factor $R$ is poor: $R > 6. 10^{-6}$, and white ones indicate where $R < 6. 10^{-6}$. The two circles mark 4.5 and 5.5 resels distances from the center of the PSF.
 {\bf bottom: } Contours and limitations of the same PSF combination for a $R =  2. 10^{-6}$ threshold (same scale as left Fig.). The two circles mark 5.5 and 8 resels from center.}
\label{champ_6e-6_gris}
\end{figure}
Studies are presently being carried out to determine the optimimum apodization tradeoff for transmission and dynamic range. Apodization will also be studied in association with coronography.


\section { Effects of aberrations  \label{aberrations}} 

As individual apertures in the array are void rectangles, only aperture positioning and dimensioning can affect the wavefront before prime focus.

The effects of aperture mispositioning on the optical path difference (OPD), which is a major challenge in other approaches of aperture synthesis, are significantly reduced here: a mispositioning $\Delta x$ of an aperture in the plane of the array leads to an OPD error 
$$
\Delta OPD = \lambda {\Delta x  \over p_x}.
$$
The smaller the pseudo period $p_x$, the  higher the impact factor of mispositioning will be on the OPD.  For a given wavefront quality, the highest positioning constraints will be put on the external elements.

Large diffractive arrays with long focal lengths have the advantage of large pseudo periods, thus providing wavefronts which are much more precise than the positioning of the elementary apertures that gave rise to them. For the long focal arrays considered here, the OPD errors on a plane wave crossing a Fresnel array can be approximated as follows (Massonnet 2003):
$$
\Delta OPD = {x\Delta x + y\Delta y \over 2f}  +  \Delta z { x^2 + y^2 \over 4f^2}
$$  
where $\Delta x$  $\Delta y$ and  $\Delta z$ are deviations of an aperture with regard to its nominal position, respectively  in the plane of the objective (x, y) and perpendicular to it (z). 

As the focal length of a Fresnel array is $f  = C^2 / 8 n  \lambda $, where $C$ is the side of the square array and  $n = k_{max}$ the number of Fresnel zones, we have:
$$ 
\Delta OPD =  {4 \lambda \over C} (k_x \Delta x + k_y \Delta y) + ...
$$
The coefficients affecting $\Delta x$ and  $\Delta z$ are very small, e.g. $10^{-3}$ and $10^{-6}$ respectively at the edge of an array 6 m in size and of focal length 2 km. By comparison the coefficient for $\Delta z$ is 2 in the case of a mirror and $n-1$ in the case of a refractive material of index $n$.

We studied the effects of two types of deformations that may occur on an array: a random "jigsaw" perturbation of the aperture edges and a "parallelogram" shear.

 For "aberrated" arrays, a 2D transmission law is generated and arbitrary errors added to the element positioning, then a complete 2D Fresnel transform is computed.

In the jigsaw deformation, the maximum amplitude of perturbations  is constant throughout the array: $\lambda/4$ peak-to-valley on the wavefront from the smallest (outer) apertures in the array. This corresponds to a cut error of $\simeq$ 0.3mm on a 6m array with 600 Fresnel zones.
\begin{figure}
 \includegraphics[width=88mm]{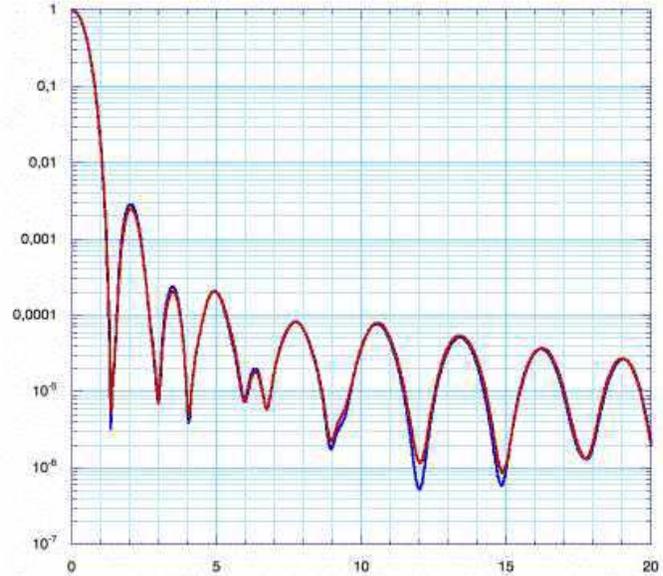} 
\caption{   Diagonal cut of the PSF of an array with  "jigsaw" perturbations on the element edges, corresponding to $\lambda/4$ PTV on the wavefront from the smallest (outer) elements. The PSF of a perfect array is drawn for comparison. These two curves closely overlap.}
\label{psf_aberee2}
\end{figure}
Due to limits in computing equipement, these results were obtained for an array with only 100 zones (center to edge) i.e. 80000 apertures.

Another possible perturbation in a Fresnel array may be caused by a "parallelogram" shear on a non-perfectly rigid frame. Numerical simulations were performed on a 100-zone array, with a corner shift of one half edge period, resulting to a shift of $\lambda/2$ for the wavefront from the most external patterns. The PSF has been computed and we see that this distortion affects the dynamic range close to the central peak  ( Figs.~\ref{image_psf_tordue2} and ~\ref{psf_tordue2} ).
\begin{figure}
\includegraphics[width=88mm]{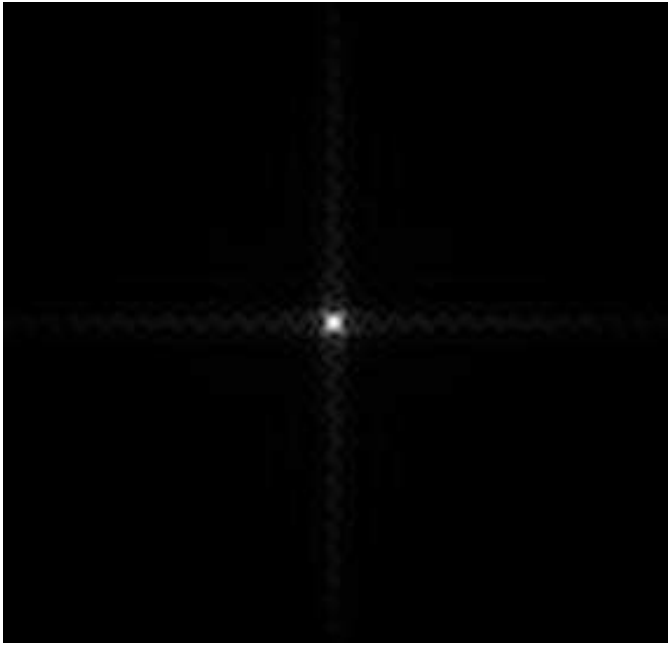}
\caption{  2-D  PSF of an array with a "parallelogram" shear distortion. The square root of the PSF intensity is displayed. Distortion effects are clearly visible on the diffraction spikes. The Strelh ratio is $\simeq$ 71\%.}
\label{image_psf_tordue2}
\end{figure}
\begin{figure}
 \includegraphics[width=88mm]{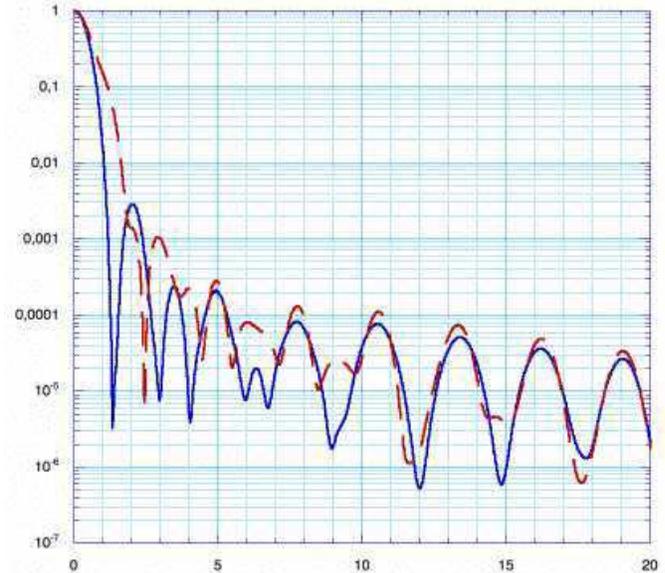}
\caption{ Diagonal cut PSF of an array with a "parallelogram" shear distortion (dashed line) compared to the PSF of a perfect array (full line). The central part of the PSF might become unusable depending on the target, but the outer peaks are only slightly brighter for the disturbed PSF.}
\vskip 1 cm
\label{psf_tordue2}
\end{figure}
%


\section { Achromatizer design and local bandwidth limitations \label{achromatizer} }  

An achromatisation scheme for holographic lenses   ( Fig.~\ref{opt_reprise} ) has been studied and published by Falklis \& Morris (1989). It uses an achromatic optical element, acting as a field lens, which reimages the primary Fresnel zone plate on a secondary one in a pupil plane. This secondary zone plate operates at order $m = -1$ and compensates for the chromatic aberration of the primary. It is  coupled to a converging achromatic element, lens or mirror, which makes the beam converge into a final focal plane.\\

For a Fresnel array, a similar design may be used, where the primary  array is achromatized using a small circular  or orthogonal zone plate. Field optics ($L_2$) in the focal setup reimage the array on this secondary zone plate (part of $L_3$), which is carved on a reflective surface and blazed for efficiency. The size of the secondary is planned to be typically  $1/200$ of the primary and may operate at order $m = -2$ rather than -1. The field relay optics  ($L_2$) consist of a Cassegrain or similar two-mirror combination.

Contrary to chromaticism corrections made with dispersive materials, this diffractive correction works for all wavelengths with no approximation. However, the bandwidth is limited by the size of the field optics.

The chromaticism correction setup also acts as a stop that blocks the other orders of diffraction, which add unfocused light and destroy the high dynamic range: all orders except $m=1$ are focused by the field optics and fall at, or close to, the center of the secondary Fresnel optics, where a mask blocks the light.

Stray light from objects out of the field, not focused by the front array, is focused by the field optics and falls on the edges of the pupil plane. As the front array and the stray light are not focused exactly in the same plane by the field optic, blocking the stray light in all cases requires a narrow opaque margin surrounding the front array. This additional band is taken into account in the formation flying study by Guidotti (2004).

 For a given position of the  prime focal plane, only one wavelength ($\lambda_0$) is strictly in focus   at that plane . At wavelengths  far from $\lambda_0$, the half width of the defocused PSF can be expressed by
$$ \rho (\lambda) = {C (\lambda - \lambda_0) \over 2 \lambda}  \,,$$
where $C$ is the size of the primary array.
 
Field optics of size $S$ at the focal plane correct the chromaticism and yield a diffraction limited PSF, but only for the light that can be captured by the field aperture $S$. For wavelengths too far from $\lambda_0$, one gets $\rho (\lambda) > S$ and vignetting effects occur, which affect both the transmission and the resolution. 

With a primary array of size $C$ and a field aperture of size $S$, the bandpass $\Delta \lambda$ centered on $\lambda_0$ is given by:
$$ \Delta \lambda = 2  \lambda_0 {S - field \over C}  $$
where $field$ is the linear non vignetted field at the focus. There is a tradeoff between field and bandpass. For example, a 6 m array with 50 cm diameter focal optics yields from a maximum bandpass at zero field: 
$${\Delta \lambda  \over \lambda_0} = {1\over 6}$$
to a maximum field $\alpha$ at null bandpass:  
$$\alpha = {8 k_{max} \lambda S \over c^2}$$
and for a $k_{max} =600$ zone array, a $field/resolution$ ratio of:
$$ S/ \rho =  8 k_{max} {S\over C} = 400$$

For $\lambda$ outside $\lambda_0 \pm \Delta \lambda/2$, the image is still achromatic, but the transmission and angular resolution decrease. At the center of the field the residual transmission is given by:
 $$ trans (\lambda) =  trans (\lambda_0 )  \left ( {S \over C} { \lambda_0 \over \lambda - \lambda_0} \right )^2  $$
\begin{figure*}
\centering
\includegraphics[width=15cm,height=5cm] {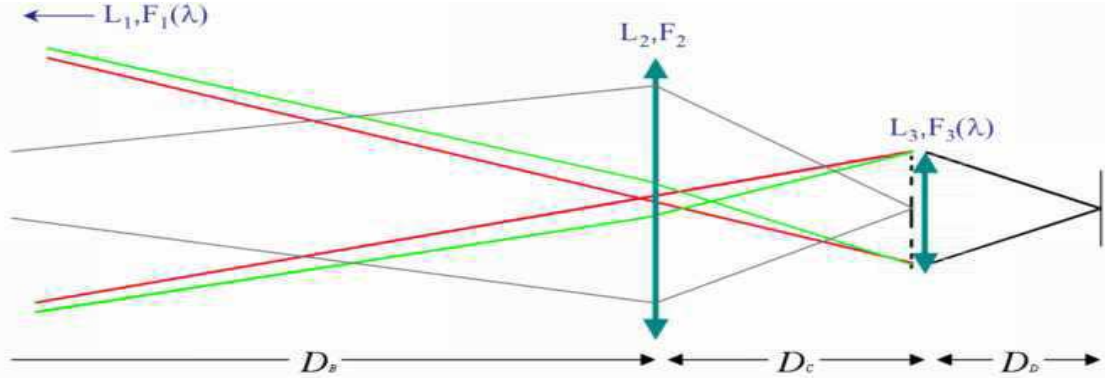}
\caption{  Achromatizer setup : 
$D_B$ is the distance from primary array ${L}_{1}$ to field lens ${L}_{2}$ placed in the primary focal plane, $D_C$ is the distance from ${L}_{2}$ to the ${L}_{3}$ set, and $D_D$ is the distance from ${L}_{3}$ set to final focal plane. Distance $D_A$, not shown on this scheme, is the distance from the target to the front lens.}
\label{opt_reprise}
\end{figure*}
As stated by Falklis and Morris (1989) ,  ${F}_{2}$ and ${F}_{3}$ must comply with the expression:\\
\\
\centerline { $\frac{1}{F_{2}} = \frac{1}{D_B} + \frac{1}{D_C}$
and
$\frac{1}{F_{3_{\lambda}}} = \frac{1}{D_C} + \frac{1}{D_D} + \frac{D_B^{2}}{D_C^{2}}*[\frac{1}{D_A} + \frac{1}{D_B} - \frac{1}{F_{1_{\lambda}}}]$}\\
\\

In the Fresnel array concept, a satellite supports the Fresnel array, another one supports the $L_2$ field lens and focal equipment. Lens $L_3$ is in fact two lenses: a Fresnel achromatising zone plate, (${L_{3_{\lambda}}}$) and an achromatic element (re-imaging part, $L_3$). For a perfectly achromatic action of $L_2$ and $L_3$, the use of mirrors rather than lenses seems appropriate. Note that $F_2$ is linked to $L_3$ dimensions, or vice versa. In order not to have a too small $2^{nd}$ Fresnel lens, i.e. a manageable size for its edge elements, a long enough $D_C$ distance can be set to be virtual by the use of a Cassegrain configuration for $L_2$. An optical combination allowing sufficient off-axis performances for re-imaging the most external patterns is presently under study.


\section { Exoplanet detection \label{detection} } 

We propose to use a Fresnel array to detect exoplanets in the "imaging" mode, i.e. using a quasi whole field rejection of the star light as described in section~\ref{dynamic}, and a high angular resolution image of the planet in that field. Other observing modes could be used with this interferometric array, such as the "Nuller" approach: any synthetic aperture setup that provides achromatic and straight null zones in its PSF is potentially usable. The Fresnel array can be designed to yield one (or two perpendicular) null zones extending from the center of the PSF, by shifting the aperture positions by a half period in symmetric halves or quadrants of the array. 

To explore direct exoplanet detection by standard imaging with a Fresnel array, we have simulated several exoplanet situations, and for each one we have chosen the smallest array size that allows detection with a ${planet \, signal / noise} > 3$ in a maximum  exposure of five hours. We used the PSF of an undistorted apodized array. We present the determined configurations and the evolution of the $planet \, signal / noise $ as a function of $\lambda$, in the spectral band for which the system is resolved angularly. The $ planet \, signal / noise $ ratio is derived from the $planet /  star$ contrast, the flux received from the star, the zodiacal and exozodiacal light contribution, the angular resolution of the array and the thermal emission from the grid forming the array. 

In the proposed instrumental setup, a baffle protecting the Fresnel grid from direct sun prevents specular reflections and keeps its temperature down to $\simeq$ 60K without active cooling. The thermal noise due to emission from the array is computed with an emissivity of  0.1. This noise is small compared to the residual star light at the planet position in all cases presented below. The baffle and the related orbital requirements at the Sun-Earth Lagrangian points $L_1$ and $L_2$ have been studied by Guidotti (2004).

\subsection { Star contribution \label{resistar} } 

The residual power from the star, integrated over  a solid angle of $(2 \lambda/C)^2$ at the planet's position in the image plane is computed as follows:
$$ B_e ={ 2 h \nu^3  \over {c^2  [\exp {h \nu \over K T_e} -1] } }$$
$$ P_e = \epsilon_e B_e  \pi \left ({R_e \over D} \right )^2  C^2  \Delta \nu  \, Trs  \, Rej $$
$T_e$, $ \epsilon_e$, $R_e$ are respectively the photosphere temperature, emissivity and stellar linear radius. The star is approximated by a uniform disc.\\
$\nu$ is the frequency corresponding to the central wavelength used,\\
 $\Delta \nu$ frequency range corresponding to the bandpass limited by the achromatizer.\\
$D$ is the distance of the stellar system, $C$ is the size of the Fresnel array,\\
$Trs$,and $Rej$ are respectively the combined transmission efficiency of the lens and optical train and the rejection factor at the planet's position.\\

\subsection { Zodiacal  and exozodiacal light contribution \label{exozo} }

The contribution of both the zodiacal and exozodiacal light are integrated over the PSF and over the line of sight. We consider the exozodiacal contribution to be the zodiacal light of a solar-system at 10 Pc. 

The local zodiacal light is sampled at $90^\circ$ from the sun, $30^\circ$ from the ecliptic plane. The exozodiacal light is considered at the same angles, corresponding to an orbit inclination angle $i=60^\circ$ and a planet at elongation. 

 We use the data in "Astrophysical quantities" (Cox \& Allen, 1999), Table 13.7  for the visible at 0.5 $\mu$m, and Fig. 13.1 and 13.2 for the I.R. at 10.9 $\mu$m. For spectral dependence, we modulate these values using a 5800 K blackbody (diffusion in the visible) plus a 275 K blackbody (emission in the I.R.)
  
The zodiacal and exozodiacal sources are considered as optically thin and extended objects. The data in Astrophysical quantities correspond to zodiacal luminance observed from within the ecliptic plane. The exozodiacal luminance integrated over the line of sight for an "asterocentric" distance  $D_p$ = 1 A.U. is twice as high. Its dependance on $D_p$ is approximated by $ I(D_p) / I( 1AU) = D_p^{-2.5}$.

Extended light sources contribute to the detection noise by their integration over all the PSF, including side lobes. In this respect, Fresnel arrays are different from other interferometric devices such as imaging or nulling setups using diluted apertures.

The PSF of a nulling interferometer tends to zero at some locations in the image field, placed at the star's position. It remains high within broad side lobes designed to cover a large part of the planetary system, where a planet is searched for, but not imaged.

An imaging interferometer using diluted apertures has broad side lobes too. The noise contribution from extended objects is affected by these side lobes, even in the case of  pupil densification.

A Fresnel array is a particular case of an imaging interferometer whose PSF is close to zero for all the field except a small region: the central peak. Its side lobes consist of narrow and dim spikes. Both the star and the planet are imaged, but the contribution from the star at the planet's position is rejected by apodization.  For the apodisation law considered here, the energy percentage in the central peak of the PSF $E_{peak}$, has been evaluated by numerical integration to $E_{peak} = 77\%$, less than but comparable to the PSF of a solid aperture. 

"Leakage" in the case of zodiacal light refers to the contribution from these extended sources integrated over the whole PSF, side lobes included. With a diluted aperture array, it corresponds to an integration over a large part of the exoplanetary system. With a Fresnel array, as in the case of a single aperture instrument, the main contribution of an extended source to a given point of the image plane is the integration over a solid angle defined by the central peak of the PSF. 

For the different exoplanets, array sizes and wavelengths considered here, the angular extension of the PSF peak varies from 0.001 to 0.2 planet orbit radius.  We have taken into account the zodiacal and exozodiacal contributions over the whole PSF, but approximated them by extended objects of uniform brightness adjusted to the star-planet distance in each case. This leads to the following expression for the total power received at focus from zodiacal light in an area corresponding to the support of the central peak of the PSF:
$$
P_{ZZ} = 4  \lambda^2 (L_{Z} + L_{EZ}  )   \Delta \lambda  \; Trs
$$
$L_{Z}$ and $L_{EZ}$ are the zodiacal and exozodiacal spectral luminances in $W m^{-2} Sr^{-1} \mu m^{-1}$\\
$Trs$ is the transmission efficiency of the lens and optical train.

\subsection { Thermal noise from the array \label{therm} } 

The thermal noise contribution from the Fresnel grid seen from the focal plane is also computed as a power:
$$ B_g = {2 h \nu^3  \over {c^2  [\exp {h \nu \over K T_g} -1] }  }$$

$$P_g = 4 \lambda^2  \epsilon_g \,B_g \,  fill \,   \Delta \nu$$

$T_g$ is the grid temperature, $\epsilon_g$ the grid emissivity; $fill$ the filling factor for the grid.

\subsection { Exoplanet contribution \label{exopla} } 

The power received from the planet at the focal plane is computed from two contributions, thermal and albedo:
$$ P_{pl} = P_{th} +P_{al} $$ 
All simulated spectra are based on the Planck emission at the considered grid, planet and star temperatures. We adapt albedos and emissivities to each situation, but do not take into account the variations of albedo or emissivity within the considered spectral bands. We do not take into account absorption in the planetary atmosphere. 

For the thermal emission $P_{th}$, we use the same expression as for the star, replacing the temperature, emissivity and radius by their planetary counterparts, $T_p$,  $ \epsilon_p$,  $R_p$.

Considering that the planet is at maximum elongation, half the disc is illuminated. The reflected part is expressed
$$ P_{al}  = 
 \epsilon_e \, B_e \,  \Delta \nu  \, \pi \left ({R_e \over D_p} \right )^2   {\pi \over 2} \, a_p \left ({R_p \over D} \right )^2  C^2 \, trs \,  E_{peak } $$
$D_p$ is the star - planet distance, $a_p$ and $R_p$ are the planet albedo and radius, respectively. \\
 $E_{peak}$ is the percentage of the energy in the central peak of the PSF\\

\subsection { Signal to noise ratio \label{snr} }

The number of detected photons from each source, star, planet, zodiacal, exozodiacal and grid, is computed from the integration through bandpass of the power received and the quantum efficiency of the detector:
$$ N_e = qe { P_e  \over h \nu }  $$

For all cases we consider\\
a 2.1\% transmission at focus for the apodized Fresnel array, \\
a 50\% efficiency for the focal optics train, including the achromatizing zone plate, \\
 70\% quantum efficiency for the detector,\\
 75\% filling factor for the grid.

The $ planet\, signal / noise $ ratio is computed assuming that the noise is the standard deviation of the number of photons detected from sources other than the planet. As the number of "noise" photons involved is in all cases higher than $10^3$, the read noise is considered negligible.

We have seen previously that for a 600-zone apodized Fresnel array, the rejection factor is\\
- better than $6 \; 10^{-6}$ in a wide part of the field situated between 4.5 and 5.5 resels from the center of the PSF,\\
- better than $6 \; 10^{-6}$ in the whole field beyond 5.5 resels,\\
- jumps to better than $2 \; 10^{-6}$ for separations larger than 8 resels, if the angular resolution leads to such a separation.\\
The threshold changes abruptly as we shift from one high point of the PSF to the next in rank, depending on the angular position in the field. As a result, the signal-to-noise curves presented in this paper show a discontinuity at wavelengths corresponding to a 8 resel planet-star separation.

We present six situations for which we have computed the required 600-zone array size for planet detection at at least 3 $\sigma$ around a solar-type star at 10 parsecs, in a maximum integration time of 10 hours (two five-hour rotated exposures). Note that for wide arrays, the number of zones could easily be raised beyond 600, leading to better rejection rates and smaller required apertures than the conservative approach taken here. These situations are

- a warm (300K)  Jupiter at 1 ua, observed with a 6 m array, 0.5 m field optics, 10 {\it minutes} of integration (Fig.~\ref{jup_chaude}).

- a cold (150K) Jupiter at 5 ua, 6 m array, 0.5 m field optics, 10 hours of integration (Fig.~\ref{jup_froide}).

- a Venus (450K) at 0.7 ua,  15 m array, 1 m field optics, 10 hours of integration (Fig.~\ref{venus}).

- an Earth (300K) at 1 ua,  40 m array, 3 m field optics, 10 hours of integration (Fig.~\ref{terre}).

- a cold (150K) Jupiter at 5 ua, same 40 m array, same 3 m field optics, 10 hours of integration, allowing detection in the I.R. (Fig.~\ref{jup_IR}).

- an Earth (300K) at 1 ua, 120 m array, 3 m field optics, 10 hours of integration, allowing detection in the I.R. (Fig.~\ref{terre_IR}).

The following curves plot the S/N ratio as a function of central wavelength, starting at $\lambda = 380 {\rm nm}$. The maximal wavelength is limited by the constraint that the star-planet separation is at least 4.5 resels. A vertical line marks the wavelength corresponding to a 5.5 resels star - planet separation.\\
\begin{figure}
\includegraphics[width=88mm]{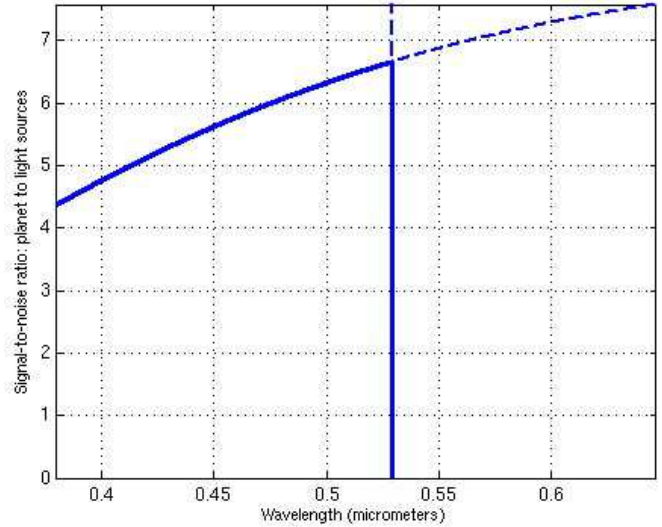} 
\caption{ Warm Jupiter at 1 AU from a solar-type star at 10 Pc, 6m array, 10 minutes integration,
wavelengths from 380nm to 650nm. Minimal S/N ratio is better than 4 at the shortest wavelengths.}
\label{jup_chaude}
\end{figure}
\begin{figure}
\includegraphics[width=88mm]{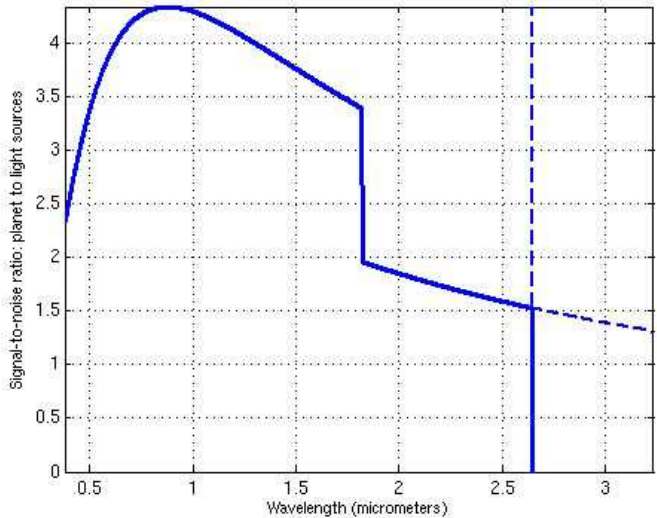}
\caption{  Cold Jupiter at 5 AU, 6m array, 10 hours integration,
wavelengths from 380nm to 3.2$\mu$m. S/N ratio is higher than 4 at $\simeq$ 900nm.}
\label{jup_froide}
\end{figure}
\begin{figure}
\includegraphics[width=88mm]{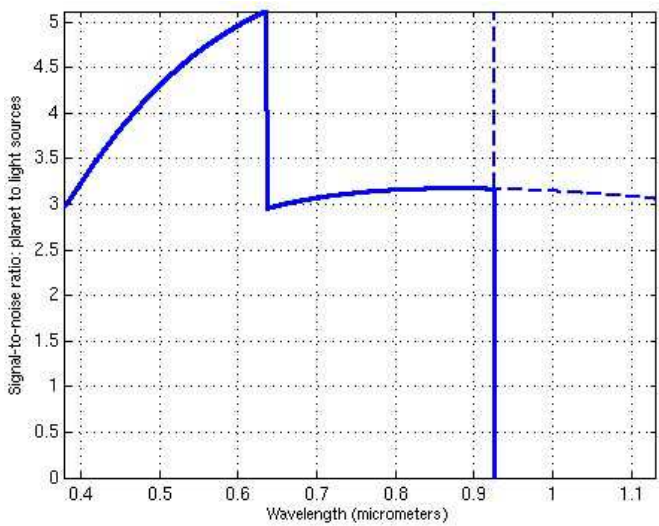} 
\caption{ Venus at 0.7 AU, 15m array, 10 hours integration,
wavelengths from 380nm to 1.1$\mu$m. S/N $\geq$ 3 throughout the observable wavelengths.}
\label{venus}
\end{figure}
\begin{figure}
\includegraphics[width=88mm]{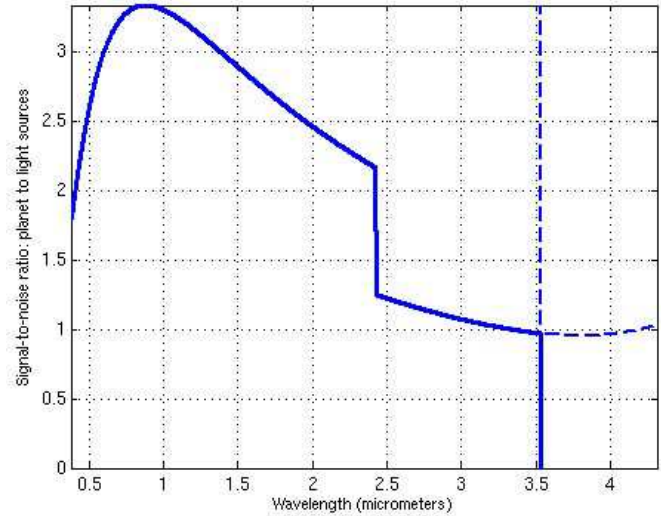} 
\caption{  Earth at 1 AU, 40m array, 10 hours integration,
wavelengths from 380nm to 4.3$\mu$m. S/N $\geq$ 3 at $\simeq$ 900nm.}
\label{terre}
\end{figure}
\begin{figure}
\includegraphics[width=88mm]{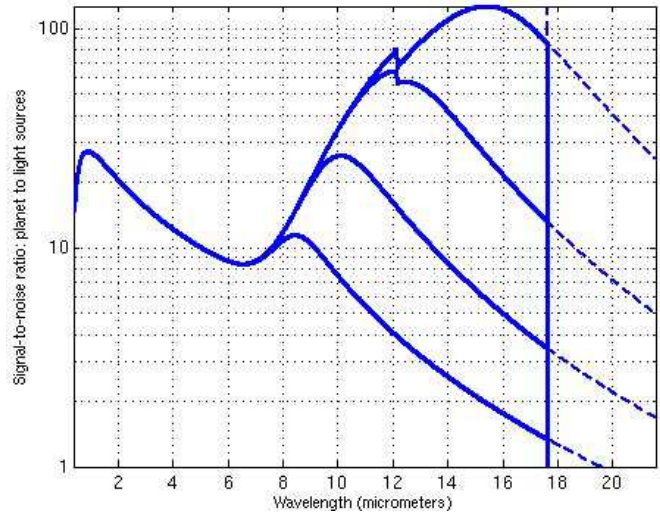} 
\caption{  Cold Jupiter at 5 AU, 40m array, 10 hours integration,
wavelengths from 380nm to 21$\mu$m. The S/N in I.R. depends upon the grid temperature: 40K to 70K, top to bottom curves, but remains $\geq$ 3 out to 17 $\mu$m below  60K. }
\label{jup_IR}
\end{figure}
\begin{figure}
\includegraphics[width=88mm]{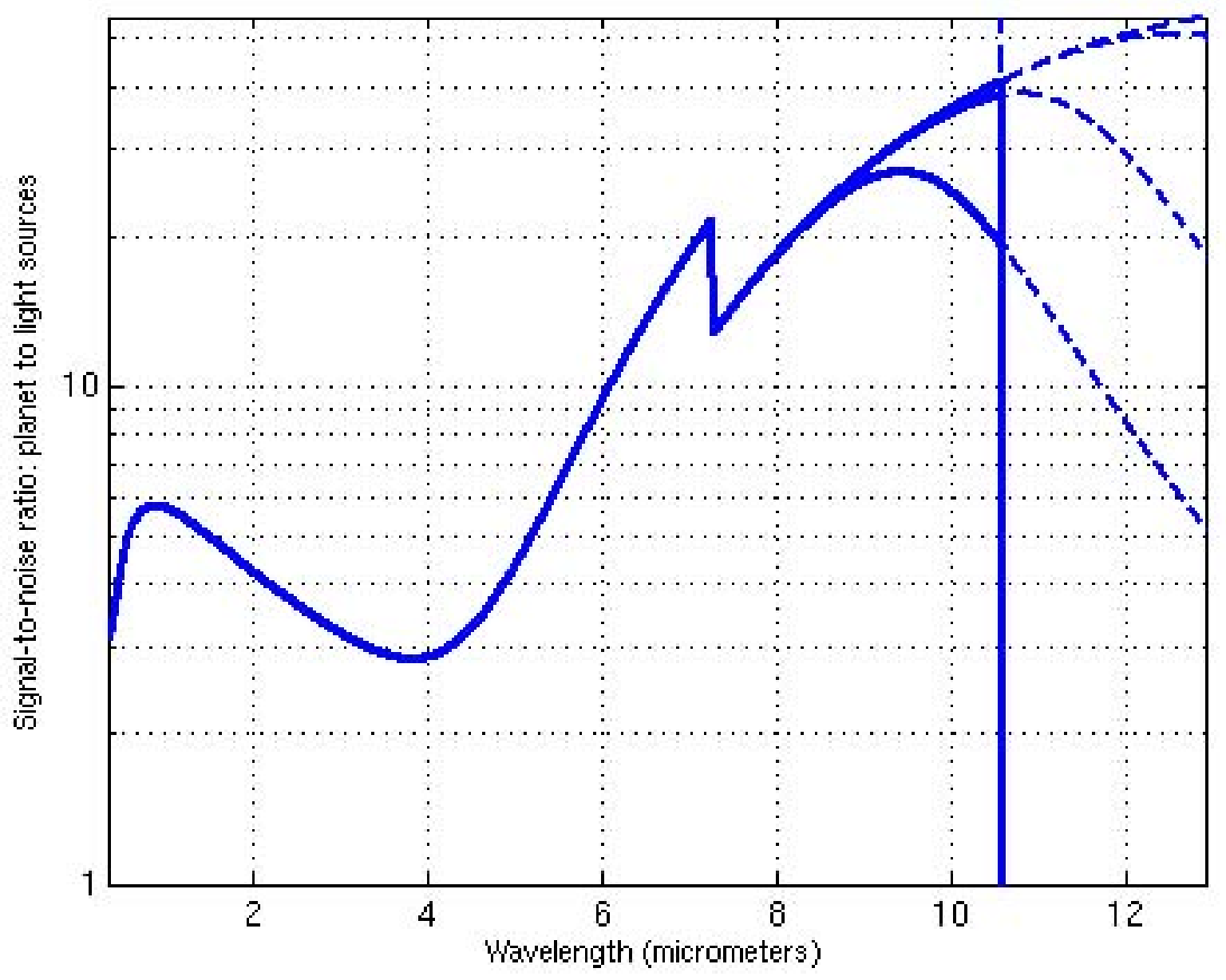} 
\caption{  Earth at 1 AU, 120m array, 10 hours integration,
wavelengths from 380nm to 12.9$\mu$m. }
\label{terre_IR}
\end{figure}
One can see that two spectral domains are favorable: one in the visible and one around 10 $\mu$m. Although $planet \over star$ ratios are less favorable in the visible, imaging exoplanets in the albedo dominant region improves detectability in three ways: first by enhancement of the rejection factor, due to a wider angular separation relative to the resolution, second by a stronger signal from the planet, and third by avoiding a dependence on the planet's temperature. Notice also that typically for cold Jupiter and Earth imaging situations (Fig.~\ref{jup_froide} $\&$ Fig.~\ref{terre}), the imaging wavelength domain ($\simeq$900nm) is far from the 8 and 5.5 resels limitations, so imaging such planetary systems at these wavelengths is possible with smaller arrays, at the cost of increased exposure time.

Limitations due to front grid emission are clearly noticeable for the standard Jupiter imaged by a 40 m array, showing different signal-to-noise evolutions beyond 8 $\mu$m, depending on the grid temperature. This shows that for imaging at wavelengths up to 18 $\mu$m, a baffle that keeps the grid temperature $\simeq$ 60K is sufficient to detect a Jupiter-like planet. 

 In our simulations, zodiacal and exozodiacal contributions to the noise remain below 1\%, except in the case of a "cold Jupiter" between 8 and 12  $\mu$m, where the zodiacal light contribution reaches a maximum of 9\% at 10  $\mu$m, and in the case of an exo-earth at 10  $\mu$m, where it reaches 4\%.

As the resolution of an array a few meters wide will not allow telluric exoplanet detection at IR wavelengths, front grid cooling does not seem to be an issue. Large arrays allow very high angular resolution and dynamic range observations even deeper in the infrared for specific targets, thus benefit from an active cooling to reduce their thermal emission.

\section { Conclusion}

In this paper, we have proposed a concept of a high resolution, high dynamic range imager that is able to detect and image exoplanets. This can be considered as an interferometer with a very large number of apertures. It can only be operated in space and requires formation flying. It also requires building precision orders of magnitude less than a reflective surface or standard interferometric array of the same angular resolution and dynamic range. The light collected over a large area is concentrated on smaller classical optics (e.g. 1/20th of the array size), to form a final image. 

This study on exoplanet detection is just one example of the applications of a Fresnel grid, which cover a large fraction of what can be done in the high angular resolution domain with a space telescope of equivalent size, such as imaging compact objects like $\eta$ Car, the envelopes of Be stars or the galactic center. As this instrument concept provides large fields ($10^6$ resels), a high angular resolution and a high dynamic range, it allows imaging of targets from the inner solar system to extragalactic objects.

Many aspects of this system remain to be optimized before a space project can be envisioned, such as the achromatizer and a better apodization scheme, for example a PIAAC at the focus (Guyon et al, 2005).

A pre-study of the flying characteristics, pointing performances in terms of speed and precision, and ergol requirements for a five-year mission at the L2 Sun-Earth Lagrangian point is available in Guidotti (2004). A breadboard project is under construction at Observatoire Midi Pyr{\'e}n{\'e}es and will serve for optical tests in the next two years.


\begin{acknowledgements}
     This work was supported by the Centre National de la Recherche Scientifique, the Universit\'e Paul Sabatier, the Fonds Social Europ\'een and Alcatel Space.\\
     We thank the anonymous referee for his/her remarks.
     \end{acknowledgements}



\begin{thebibliography}{}
%
\bibitem{Aime} Aime, C., Soummer, R., Ferrari, A., 2002,  A \& A  389, 334.
%
\bibitem{N. } Cox N. Allen C. 1999  "Astrophysical Quantities" 4th ed. (Springer) p. 147.
%
\bibitem{Falklis} Falklis D., Morris G.M., 1989,  Optical engineering, Vol 28 No 6, 592.
%
\bibitem{Guidotti}Guidotti  P.Y. , 2004, "Imageur holographique de Fresnel: vol en formation d'un syst\`eme bi-satellite" m\'emoire fin d'\'etudes, Ecole nationale sup\'erieure de l'a\'eronautique et de l'espace, Toulouse, France.
%
\bibitem{Guyon} Guyon O., 2003, A \& A  404, 379.
%
\bibitem{Guyon} Guyon O., PluzhniK E., Galigher L., Martinache F., Ridgway S.,  Woodruff R. 2005,  Ap. J. 622, 744.
%
\bibitem{lkfield} Koechlin L., Perez J.P. 2002, in ÓInterferometry for Optical Astronomy IIÓ   SPIE Proc.   4838, 411.
%
\bibitem{Koechlin2004} Koechlin L., Serre D., Duchon P. 2004,  in  "New Frontiers in Stellar Interferometry"   SPIE Proc.  5491,  1607.
%
\bibitem{carlina} Labeyrie, A. Borkowski, V., Martinache, F., Peterson, D. 2004,  "Hypertelescope imaging: from exo-planets to neutron stars",  SPIE Proc. 4838.
%
\bibitem{Lipson} Lipson S.G., Lipson H., Tannhauser D.S., 1995, Optical physics, (Cambridge University Press).
%
\bibitem{Massonnet} Massonnet D., 2003  C.N.E.S.  patent: Òun nouveau type de t\'elescope spatialÓ ref 03.13403.
%
\bibitem{Nisenson} Nisenson, P. Papaliolios  C., 2001,Ap.J., 548, L201.
%
\bibitem{Skinner} Skinner, G. von Ballmoos P., Gehrels N., Krizmanic J., 2003,   Optics   for EUV, X-ray and Gamma-Ray Astronomy", SPIE proc 5168.
%
\bibitem{Soret} J.L.Soret. 1875 Ann.Phys.Chem., 159.
%
\bibitem{Soummer} Soummer R., Aime C., Falloon P.E., 2003,  A \& A  397, 1161.
%
\end{thebibliography}
 \end {document}